\documentclass[prl,twocolumn]{revtex4}
\usepackage{amsmath}
\usepackage{amsfonts}
\begin{document}
\title{The Role of Anyonic Excitations in Fast Rotating Bose Gases
Revisited}
\author{A. \surname{Lakhoua}$^a$, M. \surname{Lassaut}$^a$, T.
\surname{Masson}$^b$, J.C. \surname{Wallet}$^b$}
%\maketitle
%\begin{center}
\affiliation{$^a$ Groupe de Physique Th\'{e}orique, Institut de Physique Nucl\'{e}aire, 
F-91406 Orsay CEDEX, France}
\author{ }
\affiliation{$^b$ Laboratoire de Physique Th\'{e}orique, B\^{a}timent 210, 91405 Orsay CEDEX, France}
%\end{center}
\begin{abstract} 
The role of anyonic excitations in fast rotating harmonically trapped Bose
gases in a fractional Quantum Hall state is examined. Standard Chern-Simons
anyons as well as "non standard" anyons obtained from a statistical
interaction having Maxwell-Chern-Simons dynamics and suitable non minimal
coupling to matter are considered. Their respective ability to stabilize
attractive Bose gases under fast rotation in the thermodynamical limit is
studied. Stability can be obtained for standard anyons while for non
standard anyons, stability requires that the range of the corresponding
statistical interaction does not exceed the typical wavelength of the
atoms.\par
\vskip 0,4 true cm

\pacs:{PACS numbers: 03.75.Lm, 73.43.-f}

\end{abstract}
\maketitle

%\strut\thispagestyle{empty}
%\pagebreak
%\setcounter{page}{1}

%tableofcontents
%listoffigures
%\listoftables
The experimental realization of Bose-Einstein Condensation (BEC) of atomic
gases \cite{BEC} has given rise to a rich variety of phenomena and motivated
numerous investigations focused on ultra-cold atomic Bose gases in rotating
harmonic traps. BEC confined to two dimensions have been created
\cite{BEC2D} and their response under rotation has been studied. Basically,
the rotation of a BEC produces vortices in the condensate
\cite{VORT1},\cite{VORT2}. When the rotation frequency increases, the BEC
state is destroyed and for sufficiently high frequency, a state
corresponding to Fractional Quantum Hall Effect (FQHE) \cite{FQE} is expected to
possibly occur \cite{F10}. In particular, when the rotation frequency is tuned to the
characteristic frequency of the harmonic confining potential in the radial
plane, FQHE states have been predicted to become possible ground states for the
system \cite{WILK}. This observation has been followed by studies focused on the
FQHE for bosons with short range interactions \cite{WORKS1}, \cite{WORKS2}. FQHE states 
involve anyonic excitations \cite{LERDA}.
Starting from the standard Chern-Simons (CS) realization for anyons
\cite{LERDA}, it has been shown in \cite{FISCHER} that in the
thermodynamical limit, a 2D harmonically trapped rotating Bose gas with
{\it{attractive}} interactions can be stabilized (in a FQHE state) thanks to
its anyonic (quasi-particle) excitations \cite{F20}. \par
The above mentioned description of anyons is not unique. Another realization has been
investigated in \cite{NOUS} where the minimal coupling of the statistical
gauge potential to matter, which is constrained to have
Maxwell-Chern-Simons (MCS) dynamics, is supplemented by a Pauli-type
coupling, as recalled below. While the effective Landau-Ginzburg (LG)
theories stemming from each anyon realization both reproduce the (so far
observed) low energy features of the Quantum Hall Fluids (QHF), slight
differences do exist between CS and MCS anyons. These latter stem from a
statistical interaction having finite range (whereas CS interaction has zero
range) and have an additional attractive mutual interaction (absent for
CS anyons). In this note, we point out that the existence of this attractive
mutual interaction may have an impact on the ability of (MCS) anyonic
excitations to stabilize an attractive Bose gas in a FQHE state. In
particular, we find that stability requires that the range of the MCS
statistical interaction does not exceed the typical wavelength for the
atoms. The description of QHF within the MCSLG theory stemming from the MCS
description of anyons is also discussed.\par
Consider an interacting Bose gas at zero temperature 
in a rotating harmonic trap with strong confinement in the direction of the 
rotation axis so that the system is actually two dimensional. The
Hamiltonian in the rotating
frame \cite{DALF} and the corresponding action quoted here for further convenience
are \cite{FOOT1}
$$H=\sum_{A=1}^N{{1}\over{2m}}({\bf{p}}_A-{\bf{A}}({\bf{x}}_A))^2
+W+V+... \eqno(1),$$
$$S_0=\int_t(\sum_{A=1}^N{{m}\over{2}}{\dot{\bf{x}}_A}^2-W-V)
+\int_{x}{\bf{A}}({\bf{x}}).{\bf{J}}({\bf{x}},t)+...\eqno(2),$$
where
${\bf{J}}({\bf{x}},t)$$=$$\sum_A$${\dot{\bf{x}}_A(t)}\delta({\bf{x}}-{\bf{x}}_A(t))$, the 
external gauge potential $A_i({\bf{x}}_A)$$=$$m\omega\epsilon_{ij}x^j_A$ yields the
Coriolis force, $W$$=$$\sum_{A=1}^N$${{m}\over{2}}(\omega_T^2-\omega^2)|{\bf{x}}_A|^2$ 
(${{\omega}\over{2\pi}}$(resp. ${{\omega_T}\over{2\pi}}$) is
the rotation (resp. trapping) frequency), $V$$=$$\sum_{A<B}V({\bf{x}}_{AB})$, $V({\bf{x}}_{AB})$ is the 
two-body potential felt by the bosons in the plane. It can be well
approximated by $V({\bf{x}}_{AB})$$=$$g_2\delta({\bf{x}}_{AB})$ since the
scattering between ultra-cold bosons is dominated by the s-wave \cite{DALF}. Here, $g_2$ can be viewed as an effective coupling constant
encoding basically the effects of the harmonic axial confinement (with trapping frequency ${{\omega_z}\over{2\pi}}$ and localization
length $l_z$$=$${\sqrt{{{1}\over{m\omega_z}}}}$) along the rotation axis on the initial 3D scattering properties of the
atoms \cite{GANG}. This will be discussed later on. In (1)-(2), the ellipses denote possible
multi-body interaction terms. We now consider the "limit"
$\omega\simeq\omega_T$ for which the trapping and centrifugal potential
(nearly) balance each other so that $W$ can be
neglected.\par
The standard CS description of anyons \cite{LERDA} obtained from (1)-(2) is achieved by
minimally coupling to $S_0$ a statistical CS gauge potential $a_\mu$:
$S_0$$\to$$S_1$$=$$S_0$$+$$\int_x{{\eta}\over{4}}\epsilon_{\mu\nu\rho}a^\mu
f^{\nu\rho}$$+$$a_\mu J^\mu$ where the CS parameter $\eta$ is dimensionless,
$J_\mu$$=$$(\rho;{\bf{J}})$ with
$\rho$$=$$\rho({\bf{x}},t)$$=$$\sum_A$$\delta({\bf{x}}$$-$${\bf{x}}_A(t))$ and ${\bf{J}}$ as given above
($f_\mu$$\equiv$${{1}\over{2}}\epsilon_{\mu\nu\rho}f^{\nu\rho}$, 
$f_{\mu\nu}$$=$$\partial_\mu a_\nu$$-$$\partial_\nu a_\mu$, $f_0$ (resp. $f_i$)
is the statistical magnetic (resp. electric) field). The equations of motion
for $a_\mu$ take the field-current identity form
$f_\mu$$=$$-$${{1}\over{\eta}}J_\mu$ (i), insuring that particle-statistical magnetic
flux composite anyonic objects are formed \cite{LERDA}. Particle-flux coupling responsible
for the  formation of the above composite quasi-particles leads to the
occurrence of Aharonov-Bohm type interactions. The anyonic
character of the wave functions for the quasi-particles is determined by
the Aharonov-Bohm phase \cite{LERDA} $\exp(i\int_C{\bf{a}}.d{\bf{x}})$$=$$\exp
{{i}\over{\eta}}$ ($C$
is some closed curve) that is induced when one quasi-particle moves
adiabatically around another one, equivalent to 
a {\it{double}} interchange of identical quasi-particles in the wave
function. Then, the statistics of the quasi-particles is controlled by the
value of $\eta$: they have Fermi(resp. Bose) statistics
whenever $\eta$$=$${{1}\over{(2k+1)2\pi}}$ (resp.${{1}\over{(2k)2\pi}}$),
$k$$\in$$\mathbb{Z}$. The statistics is anyonic otherwise. Notice that, when $\eta$$=$
${{1}\over{(2k+1)2\pi}}$, the initial Bose statistics of the atoms is
converted into a Fermi statistics for the quasi-particle which corresponds to
the so called statistical transmutation \cite{LERDA}. This, plugged into a second quantized formalism,
leads to the usual CSLG action \cite{ZHANG} underlying  the main part of the analysis 
presented in \cite{FISCHER}. Since the statistical interaction is mediated by
a CS gauge potential, its range is zero. \par
An alternative realization for anyons has been proposed and discussed in
\cite{NOUS}. It is obtained by coupling a MCS
statistical gauge potential $a_\mu$ to $S_0$ through minimal and non minimal
(Pauli-type) coupling term, the strength of this latter being fixed to a
specific value \cite{NOUS}, namely: $S_0$$\to$$S_2$
$=$$S_0$$+$$\int_x$$-$${{1}\over{4e^2}}f_{\mu\nu}f^{\mu\nu}$$+$${{\eta}\over{4}}\epsilon_{\mu\nu\rho}a^\mu
f^{\nu\rho}$$+$$a_\mu J^\mu$$-$${{1}\over{\eta e^2}}f_\mu J^\mu$ ($e^2$ has mass dimension $1$), where 
the coupling constant of the Pauli-term (last term in $S_2$) 
has been already fixed to the above mentioned specific value. Then, the equations of 
motion for $a_\mu$ can be written as
$-$${{1}\over{e^2}}\epsilon_{\alpha\mu\rho}\partial^\mu(f^\rho+{{1}\over{\eta}}J^\rho)
$$+$$\eta(f_\alpha+{{1}\over{\eta}}J_\alpha)$$=0$ which is solved by (i). This 
latter observation has been used as a starting point to construct an
effective theory reproducing the usual
anyonic behavior \cite{NOUS}. The Aharonov-Bohm phase defining the
statistics of the resulting quasi-particles still verifies $\exp(i\int_C{\bf{a}}.d{\bf{x}})$$=$$\exp
{{i}\over{\eta}}$ so that $\eta$ again controls the statistics.
However, anyons obtained through this construction have an additional attractive
contact mutual interaction (not present in the CS case) \cite{NOUS}. Combining $S_2$
with the second quantization machinery, one obtains the corresponding MCSLG action
$$S=\int_x 
i\phi^\dag{\cal{D}}_0\phi-{{1}\over{2m}}|{\cal{D}}_i\phi|^2-g_2(\phi^\dag\phi)^2
-U(\phi)$$
$$-{{1}\over{4e^2}}f_{\mu\nu}f^{\mu\nu}+{{\eta}\over{4}}\epsilon_{\mu\nu\rho}a^\mu
f^{\nu\rho} \eqno(3),$$
$${\cal{D}}_0=\partial_0-ia_0+{{if_0}\over{\eta e^2}}\ ;\
{\cal{D}}_i=\partial_i-i(a_i+A_i)+{{if_i}\over{\eta e^2}}
\eqno(4a;b).$$
In (3), $\phi$($=$$\phi({\bf{x}},t)$) is the order parameter, $\phi^\dag\phi$$=$$\rho$, $U(\phi)$
denotes the LG potential for multi-body
interactions, $U(\phi)$$=$$g_3|\phi|^6$$+$$...$ ($g_3$$>$$0$ as in
\cite{FISCHER}). The non minimal coupling 
terms have been included in the extended covariant
derivative (4). The statistical MCS gauge potential has a finite mass
\cite{DESER} $M=\vert\eta\vert e^2$, so that the statistical interaction 
has a finite range $\Lambda_{st}$$=$${{{1}\over M}}$. The
CSLG action \cite{ZHANG} is obtained from (3)-(4) by taking the limit
$e^2$$\to$$\infty$. While CSLG action  is believed to encode
the (so far observed) low energy physics of QHF, it
appears that MCSLG action also reaches this goal and could
therefore be used as an alternative description of QHF. This is
discussed more closely at the end of this letter. For the moment, we assume that 
the system is in a FQHE state ($\eta$$\ne$$0$) described
the MCSLG action (3)-(4).\par
The equations of motion derived from (3) are
$$i{\cal{D}}_0\phi+{{1}\over{2m}}{\cal{D}}_i{\cal{D}}_i\phi-2g_2(\phi^\dag\phi)\phi=
{{\delta U(\phi)}\over{\delta\phi^\dag}} \eqno(5a),$$ 
$$-{{1}\over{e^2}}\epsilon_{\alpha\mu\rho}\partial^\mu(f^\rho+{{1}\over{\eta}}{\cal{J}}^\rho)
+(f_\alpha+{{1}\over{\eta}}{\cal{J}}_\alpha)=0
\eqno(5b),$$
where ${\cal{J}}_0$$=$$\rho$, ${\cal{J}}_i$$=$
${{i}\over{2m}}(\phi^\dag{\cal{D}}_i\phi-({\cal{D}}_i\phi)^\dag\phi)$
and ${\cal{D}}_\mu$ is defined in (4) while anyonic configurations, on which
we now focus, are obtained from (5b) when 
$f_\mu$$=$$-$${{1}\over{\eta}}{\cal{J}}_\mu$ \cite{FOOOT2}. From the field's conjugate momenta
$\Pi_{\phi^\dag}$$=$$\Pi_{a_0}=0$, $\Pi_{\phi}$$=$$i\phi^\dag$,
$\Pi_{a_i}$$=$${{1}\over{e^2}}f_{0i}$$-$${{\eta}\over{2}}\epsilon_{ij}a^j$$+$
${{1}\over{\eta e^2}}\epsilon_{ij}{\cal{J}}^j$, one obtains the Hamiltonian
$$H=\int_{\bf{x}}
{{1}\over{2e^2}}(f_0+{{1}\over{\eta}}\rho)^2
+{{1}\over{2e^2}}\Theta f_i^2$$
$$+{{1}\over{2m}}\vert
D_i\phi\vert^2+(g_2-{{1}\over{2\eta^2e^2}})\rho^2+U(\phi) \eqno(6),$$
where $D_i$$=$$\partial_i$$-$$i(a_i+A_i)$ and we 
defined $\Theta$$=$$1$$-$${{\rho}\over{m\eta^2e^2}}$. The positivity of the model 
requires $\rho$$\le$$ m\eta^2e^2$ \cite{LATINSKY}. Restoring $\hbar$
and $c$, this translates 
into $\rho$$\le$$\rho_{lim}$, $\rho_{lim}$$=$
$\vert\eta\vert({{\Lambda}\over{\Lambda_{st}}}){{1}\over{\Lambda^2}}$
where $\Lambda$$=$${{\hbar}\over{mc}}$ is a typical (de Broglie) wavelength for
the atoms and $\Lambda_{st}$$=$${{\hbar}\over{|\eta|{\tilde{\mu}} c}}$
($e^2$$=$${\tilde{\mu}}c^2$). This condition should be fulfilled by current experimental values,
provided $\Lambda_{st}$$\lesssim$${\cal{O}}(\Lambda)$, a condition that we now
assume \cite{F2}. Namely, for $^7$Li, $^{23}$Na, $^{87}$Rb, one obtains respectively 
$\rho_{lim}$$\simeq$$\vert\eta\vert({{\Lambda}\over{\Lambda_{st}}})$
$10^{29}$, $10^{30}$, $10^{31}$ cm$^{-2}$, which, for possibly reachable
current Quantum Hall states, should be larger by several
orders of magnitude than the experimental values reached by the matter
density so that ${{\rho}\over{m\eta^2e^2}}$$\ll$$1$.\par
Compared to the CSLG Hamiltonian \cite{FISCHER}, $H$ involves additional
contributions from the Maxwell part of
(3) and the non minimal coupling terms in (4). The interaction energy
$g_2\rho^2$ receives contributions from ${{1}\over{2e^2}}f_0^2$ and the Pauli-type coupling
in (4a): these latter combine to yield the first term in (6) together with the
additional attractive (magnetic) contribution $-$${{\rho^2}\over{2\eta^2e^2}}$ to 
$g_2\rho^2$. This, depending on the relative magnitude of $\Lambda$ and
$\Lambda_{st}$, may somehow alter the conclusion obtained in \cite{FISCHER}
from a CSLG
description of anyons about the ability of statistical interaction to
stabilize an attractive Bose gas, as we now show.\par  
We set
$\phi$$=$$\gamma e^{i\varphi}$$\equiv$${\sqrt{\rho}}e^{i\varphi}$, ${\hat{a}}_i$$=$
$\partial_i\varphi$$-$$(a_i$$+$$A_i)$, $\eta$$>$$0$. For
$N$ atoms in the trap, one has $\int_{\bf{x}}\phi^\dag\phi$$=$$N$. Using 
$f_\mu$$=$$-$${{1}\over{\eta}}{\cal{J}}_\mu$ and assuming that fields vanish at infinity so that boundary terms
disappear, the static energy stemming from (6) can be conveniently written as
$$H=\omega N+\int_{\bf{x}}{{1}\over{2m}}(\partial_i\gamma
-\epsilon_{ij}{{{\hat{a}}^j\gamma}\over{{\sqrt{\Theta}}}})^2
+(g_2-g_0)\rho^2$$
$$+\int_{\bf{x}}m\eta^2e^2\sum_{k=3}^\infty(\eta e^2C_{k-1}+2\omega
C_k)({{\rho}\over{m\eta^2e^2}})^k
+U(\gamma) \eqno(7).$$
In (7), $\omega$$>$$0$, the positive constants $C_k$'s are given by
$(1$$-$${{\rho}\over{m\eta^2e^2}})^{{{1}\over{2}}}$$=$$1$$-$
$\sum_{k=1}^\infty C_k({{\rho}\over{m\eta^2e^2}})^k$
($C_k$$=$${{\Gamma(k-{{1}\over{2}})}\over{2k!{\sqrt{\pi}}}}$ where $\Gamma$
is the Euler function) and we
have defined
$$g_0=-{{\hbar^2}\over{2m\eta}}(1-{{\Lambda_{st}}\over{\Lambda}}
(1-{{\hbar\omega}\over{2mc^2}}))\simeq 
-{{\hbar^2}\over{2m\eta}}(1-{{\Lambda_{st}}\over{\Lambda}})\eqno(8),$$
where $\hbar$ and $c$ have been reinstalled and the rightmost relation
stems from $\hbar\omega$$\ll$$mc^2$ which holds for current experimental 
values for $\omega$ (taking ${{\omega}\over{2\pi}}$$\sim$${\cal{O}}(10)$-${\cal{O}}(10^3)$ Hz as a
benchmark). In the same way, the term $\sim$$\omega C_k$ in (7) can also been neglected
(since $\hbar\omega$$\ll$$\eta{\tilde{\mu}} c^2$ in view of
$\Lambda_{st}$$\lesssim$${\cal{O}}(\Lambda)$).\par
The quartic interaction terms in (7) can be eliminated provided $g_2$ is
chosen to be
$$g_2=g_0 \eqno(9).$$
Then, neglecting for the moment in (7) the small terms of order $\sim$${\cal{O}}(\gamma^6)$ (and
higher) as in \cite{FISCHER}, the ground state of (7) is obtained for those
configurations satisfying
$(\partial_i\gamma$$-$$\epsilon_{ij}{{{\hat{a}}^j\gamma}\over{\Theta^{{{1}\over{2}}}}})$$=$$0$,
$i$$=$$1,2$. This, combined with (5a), (8), (9), further expanding the various
contributions depending on $\Theta$ in powers of ${{\rho}\over{m\eta^2e^2}}$ and using the fact that
${{\rho}\over{m\eta^2e^2}}$$\ll$$1$ yields
$\partial_i^2\gamma$$-$${{(\partial_i\gamma)^2}\over{\gamma}}$$=$$-$$\gamma(
{{\gamma^2}\over{\eta}}(1$$+$${{\omega}\over{\eta e^2}})$$+$$2m\omega)$
where, in the RHS, ${{\omega}\over{\eta e^2}}$$\ll$$1$ still holds so that
this latter equation can be accurately approximated by
$$\partial_i^2\gamma-{{(\partial_i\gamma)^2}\over{\gamma}}=-\gamma(
{{\gamma^2}\over{\eta}}+2m\omega) \eqno(10).$$
When $\omega$ goes to zero, the
solutions of (10) reduces smoothly \cite{EZAWA} to the solutions of the Liouville equation
$$\Delta\ln\rho=-{{2}\over{\eta}}\rho \eqno(11).$$
These latter can be parametrized as
$\rho(z)$$=$$4\eta|h^\prime(z)|^2(1+|h(z)|^2)^{-2}$ ($z$$=$$x_1$$+$$ix_2$)
for any holomorphic function $h(z)$. The particular choice
$h(z)$$=$$(z_0/z)^n$ ($z$$=$$re^{i\theta}$), $n\in \mathbb{N}$, gives rise
to radially symmetric vortex-type solutions $\rho(r)$$=$${{4\eta
n^2}\over{r^2}}(({{r_0}\over{r}})^n+({{r}\over{r_0}})^n)^{-2}$ where $r_0$
is some arbitrary length scale and $\phi$$=$${\sqrt{\rho}}e^{i n\theta}$.\par
Let us discuss the above analysis. From (7)-(9), one realizes that the
initial two-body coupling term $\sim$$g_2$ can be compensated by the
statistical interaction.  In view of (7), for fixed $\eta$$>$$0$, 
the particular value $g_0$ (8) represents the limiting value for the two-body
coupling constant $g_2$ below which the Bose gas cannot be stabilized by the
statistical interaction. In the CS case, $\Lambda_{st}$$=$$0$ and $g_0$ is
negative. Then, when $g_2$$=$$g_0$, corresponding to an initial attractive
Bose gas, the system behaves as a free anyon gas whose ground state (neglecting the small $|\phi|^6$
and higher order terms in $U(\phi)$) is exactly described by (10) giving rise to non singular finite energy 
matter distribution \cite{EZAWA}. Then, one would conclude as in
\cite{FISCHER} that an
attractive Bose gas may be stabilized against collapse in the
thermodynamical limit by anyonic excitations stemming from a CS (zero range) statistical interaction. This
conclusion is somehow altered in the case of "non standard" MCS anyons.
In this case, $g_0$ receives an additional positive
contribution from the (statistical) magnetic energy and the
magnetic coupling in (4a) and the sign
of $g_0$ depends now on the relative magnitude of $\Lambda$ and
$\Lambda_{st}$ (see (8)). When
$g_2$$=$$g_0$, the system behaves again as a free MCS anyon gas whose ground
state (neglecting again small $\gamma^6$ and higher order terms) is now
accurately described by (10). However, this situation can be reached by an
initial attractive Bose gas only if $\Lambda_{st}$$<$$\Lambda$ corresponding
to negative values for $g_0$. A MCS statistical interaction with range
exceeding the typical wavelength $\Lambda$ for the atoms could not
protect an attractive Bose gas from collapse (in the thermodynamical
limit).\par
Now, let us assume that $\Lambda_{st}$$<$$\Lambda$ and discuss more closely (9) from a physical viewpoint. In (most of) current
experiments, 2D atomic systems are obtained by tightening the axial confinement applied to the initial (harmonically trapped)
3D system, which restricts the dynamics of the atoms along the axial direction to 
zero point oscillations \cite{GANG}. Kinematically, the system becomes 2D while the effective coupling constant
$g_2$ for the interparticle interaction depends closely on the motion of the atoms in the axial direction \cite{GANG}. As shown 
in \cite{GANG}, $g_2$$=$${{4\pi\hbar^2}\over{m}}$$({\sqrt{2\pi}}({{l_z}\over{a_s}})$$+$$\ln({{B}\over{\pi}}
{{\hbar\omega_z}\over{\epsilon}}))^{-1}$ ($B$$\simeq$$0.915$, $\epsilon$ is the typical energy for the relative motion of the atoms and 
$a_s$ is the (3D) s-wave scattering length) where
$\epsilon$$\ll$$\hbar\omega_z$ must hold. This regime (called quasi-2D regime in \cite{GANG}), relevant here, seems to  be accessible to 
experiments \cite{BEC2D}, \cite{GANG}. Then, (9) holds provided $a_s$ is tuned to the negative value $a_s$$=$$-$
${\sqrt{2\pi}}l_z$$({{8\pi\eta}\over{1-{{\Lambda_{st}}\over{\Lambda}}}}$$+$$\ln({{B}\over{\pi}}{{\hbar\omega_z}\over{\epsilon}}))^{-1}$. 
Typical values obtained for 
${{\hbar\omega_z}\over{\epsilon}}$$\sim$${\cal{O}}(10^2$$-$$10^3)$ are ${{|a_s|}\over{l_z}}$$\sim$${\cal{O}}(10^{-1})$ (resp.
${{|a_s|}\over{l_z}}$$\sim$${\cal{O}}(10^{-1}$$-$$10^{-2})$) for a CS (resp. MCS) statistical interaction (assuming possibly reachable 
current Quantum Hall states and $\Lambda_{st}$$\lesssim$${{1}\over{2}}\Lambda$). Let us discuss the possible effect 
of the inclusion of the next to leading order interaction terms $\sim$$|\phi|^6$ 
on the stability of the system \cite{DENSPROF}, assuming that (9) holds. In the CS case, stability requires $g_3$$>$$0$, as indicated in \cite{FISCHER}. 
In the MCS case, the $|\phi|^6$ term in $U(\phi)$ receives an additional {\it{positive}} (repulsive) 
contribution coming from the infinite sum in (7). Then, stability in the MCS case is obtained provided $g_3$$+$${{1}\over{8m^2\eta^3e^2}}$
$>$$0$. The actual computation of $g_3$, which here must correspond to the quasi-2D regime, is still lacking and 
would need to extend the analysis reported in \cite{GANG} to the case of higher order interactions within the quasi-2D regime.\par 
QHF, usually described
in the low energy regime by a CSLG theory \cite{ZHANG}, can alternatively be described by a MCSLG theory as 
defined in (3)-(4). To compare
to the CS case \cite{ZHANG}, it is convenient to add a chemical potential
term to (3), $S^{\prime}$$=$$S$$+$$\int_x$$\mu\phi^\dag\phi$ ($\mu$$>$$0$) and set $U(\phi)$$=$$0$ 
in all. (5a) becomes
$i{\cal{D}}_0\phi$$+$${{1}\over{2m}}{\cal{D}}_i{\cal{D}}_i\phi$$-$$2g_2(\phi^\dag\phi)\phi$$=$
$\mu\phi$ while the relevant anyonic configurations are still obtained
when $f_\mu$$=$$-$${{1}\over{\eta}}{\cal{J}}_\mu$ solving (5b). Then, the
equations of motion admit the uniform (constant) density solution
minimizing the energy $H^\prime$$=$$H$$-$$\int_{\bf{x}}\mu\rho$, given by
$$\phi={\sqrt{n_0}};\ {\bf{a}}+{\bf{A}}=0;\ a_0=0;\ 
n_0={{\mu}\over{{2\hat{g_2}}}} \eqno(12a;b;c;d),$$ 
where ${\hat{g_2}}$$=$$g_2-{{1}\over{2\eta^2e^2}}$, similar to the
usual uniform density solution supported by the CSLG
action for QHF\cite{ZHANG}. (12b) implies that the
external magnetic field is screened by the statistical magnetic field. This
combined with $f_0$$=$$-$${{1}\over{\eta}}\rho$, yields $\nu$$=$$2\pi\eta$
where $\nu$ is the filling factor. When statistical
transmutation occurs as expected in Hall systems,
$\nu$$=$${{1}\over{2k+1}}$, $k$$\ge$$0$. Introducing
polar coordinates $(r,\theta)$ and setting
$\phi$$=$${\sqrt{\rho}}e^{in\theta}$,
${\bf{a}}$$=$${\bf{e}}_\theta{{\alpha(r)}\over{r}}$, $a_0$$=$$a_0(r)$ 
(${\bf{e}}_\theta$$=$$(\sin\theta,$$-$$\cos\theta)$, $n$$\in$$\mathbb{Z}$ is the winding number), one finds that $H^\prime$ admits 
static finite energy vortex solutions satisfying the
boundary conditions $\rho$$\sim$$n_0$, $\alpha(r)$$\sim$$m\omega r^2$ for
$r$$\to$$\infty$, $\rho$$\to$$0$, $\alpha(r)$$\to$$-$$n$ for $r$$\to$$0$.
The vortex effective magnetic flux
$\Phi$$\equiv$$\int_{{\bf{x}}}$$(f_0$$+$$F_0)$ is then $\Phi$$=$$-2\pi n$
($F_0$$=$$\epsilon_{ij}\partial^iA^j$). Using $n_0$$=$$\eta F_0$, $f_0$$=$$-$${{\rho}\over{\eta}}$, $\Phi$ 
is related formally to the "vortex
charge" $Q$$=$$\int_{\bf{x}}$$(\rho$$-$$n_0)$ through
$Q$$=$$-$$\eta\Phi$$=$$2\pi\eta n$. When statistical transmutation occurs, 
$Q$$=$${{n}\over{2k+1}}$ so that the  vortices can be interpreted as (the analogue of) 
the fractionally charged Laughlin quasi-particles, as
it is the case in the CS description of QHF \cite{ZHANG}. Finally, by expanding
$S^\prime$ around the (mean field) uniform solution (12) up to the quadratic
order in the fields fluctuations, fixing the gauge freedom, integrating
out the fluctuations of the statistical field, one obtains the low energy
effective action for the matter degree of freedom. It yields the following
dispersion relation
$$\omega_0^2({\bf{p}})=({{{\bf{p}}^4}\over{4m^2}}+{\bf{p}}^2{{\mu}\over{m}}+{{n_0^2}\over{m^2\eta^2}})(1-2\Theta_0^2)
+{\bf{p}}^2(\Theta_0+2\Theta_0^2)
 \eqno(13),$$
$\Theta_0$$=$${{n_0}\over{m\eta^2e^2}}$ ($\omega_0^2({\bf{p}})$$\simeq$
${{{\bf{p}}^4}\over{4m^2}}$$+$${\bf{p}}^2{{\mu}\over{m}}$
$+$${{n_0^2}\over{m^2\eta^2}}$ since
$\Theta_0$$\ll$$1$). This indicates that the system has
a gap, so that the
fluid described by $S^\prime$ is incompressible.\par
Although CS and MCS anyons share many similar properties (in particular those related to the statistical phase), MCS anyons 
feel an additional attractive contact mutual interaction. Therefore (when $\Lambda_{st}$$\ne$$0$), CS and MCS anyons
must be considered as 
distinct composite objects, both carrying anyonic statistics. This can be understood from the
mechanism \cite{NOUS} responsible for the formation of each type of these particle-flux anyonic composite 
(quasi-particles), reflecting itself in each case into the origin of the field-current identity i). Recall that i) 
physically expresses the fact that for a distribution of point-like particles only localized (statistical) fields can appear.
This is automatically fulfilled in the CS case (see $S_1$) since the CS statistical interaction has zero range. In the 
MCS case for which the interaction has finite range, i) results from 
an exact cancellation between electric and Pauli-type coupling effects \cite{NOUS}, which amounts to fix 
the Pauli coupling constant to a specific value (see \cite{REMARK}). This leads to the appearance
of an additional attractive contact interaction (whose strength grows as 
$\Lambda_{st}$ increases) among the resulting MCS anyons 
which, apart from this, behave essentially 
as CS anyons. The relevant question is now to examine if
some effects specific to the MCS framework, in particular those related to the above contact attraction, may be experimentally observed. 
Clearly, the ``electronic'' Quantum Hall systems are excluded: the fact that the repulsive Coulomb interaction 
among electrons cannot be manipulated combined 
with the specific features of the experimentally accessible observables make these systems only sensitive to the properties both 
shared by CS and MCS anyons in the long wavelength limit. In fast rotating (harmonically trapped) Bose gases, the interaction among
atoms can be manipulated, offering a way to study the stabilization of initially attractive Bose gases in a FQHE state through 
their anyonic excitations. As we have shown, stability may be in particular somehow conditioned by the actual nature of the 
statistical interaction. Basically, anyons of MCS origin loose their ability to stabilize the system as the range of the 
MCS interaction grows. We note that an interesting proposal to measure the statistical phase of anyons has been presented in 
\cite{PAREDES}. This, combined with an experimental implementation of the analysis presented here may provide a deeper insight into the
physical features of the anyonic composite quasi-particles.


\begin{thebibliography}{30}
\bibitem{BEC}: M.H. Anderson et al., Science 269 (1995) 198; K.B. Davis et
al., Phys. Rev. Lett. 75 (1995) 3969; W. Ketterle, Physics Today 52 (2000) 30.
\bibitem{BEC2D}: A. G\"orlitz et al., Phys. Rev. Lett. 87 (2001) 130402.
\bibitem{VORT1}: M.R. Matthews et al., Phys. Rev. Lett. 83 (1999) 2498.
\bibitem{VORT2}: K.W. Madison, F. Chevy, W. Wohlleben, J. Dalibard, Phys.
Rev. Lett. 84 (2000) 806; F. Chevy, K.W. Madison, J. Dalibard, Phys. Rev.
Lett. 85 (2000) 2223; J.R. Abo-Shaeer, C. Raman, J.M. Vogels, W. Ketterle,
Science 292 (2001) 476.
\bibitem{FQE}: For a review on FQHE see e.g. "The Quantum Hall Effect", R.E.
Prange, S.M. Girvin eds., Springer-Verlag, Berlin 1990 and references therein.
\bibitem{F10}: When the trapping and centrifugal potential balance each
other, the bosons feel only the Coriolis force. Then, the system is formally
equivalent to that of 2D (bosonic) particles in a magnetic field,
corresponding to the condition for occurrence of Quantum Hall regime.
\bibitem{WILK}: N.K. Wilkin, J.M.F. Gunn, R.A. Smith, Phys. Rev. Lett. 80
(1998) 2265; N.R. Cooper, N.K. Wilkin, Phys. Rev. B60 (1999) R16279; N.K.
Wilkin, J.M.F. Gunn, Phys. Rev. Lett. 84 (2000) 6; N.R. Cooper, N.K.
Wilkin, J.M.F. Gunn, Phys. Rev. Lett. 87 (2001) 120405.
\bibitem{WORKS1}: A.D. Jackson et al., Phys. Rev. Lett. 86 (2001) 945; B.
Paredes, P. Fedichev, J.I. Cirac, P. Zoller, Phys. Rev. Lett. 87 (2001)
010402; T.-L. Ho, Phys. Rev. Lett. 87 (2001) 060403; U.R. Fischer, G. Baym,
Phys. Rev. Lett. 90 (2003) 140402.
\bibitem{WORKS2}: M. Manninen, S. Viefers, M. Koskinen, S.M. Reimann, Phys.
Rev. B64 (2001) 245322; J. Sinova, C.B. Hanna, A.H. MacDonald, Phys. Rev.
Lett. 89 (2002) 030403; V. Schweikhard et al., Phys. Rev. Lett. 92 (2004)
040404; N. Regnault, T. Jolicoeur, Phys. Rev. Lett. 91 (2003) 030402, Phys.
Rev. B69 (2004) 235309.
\bibitem{LERDA}: For reviews, see e.g. A. Lerda, "Anyons: Quantum Mechanics
of Particles with Fractional Statistics", Lecture Notes in Physics, vol.14
(Springer Verlag, Berlin, 1992); A. Khare, "Fractional Statistics and
Quantum Theory" (World Scientific, Singapore, 1997).
\bibitem{FISCHER}: U.R. Fischer, Phys. Rev. Lett. 93 (2004) 160403; see also 
U. Fischer, Phys. Rev. Lett. 94 (2005) 208904.
\bibitem{F20}: In the thermodynamical limit, a system of attractive
untrapped bosons is unstable against collapse and is stable only for a
finite number of trapped particles; see P. Nozi\`eres, D. Saint James, J.
Phys. (Paris) 43 (1982) 1133.
\bibitem{NOUS}: J. Stern, Phys. Lett. B265 (1991) 119; I.I. Kogan, Phys.
Lett. B 262 (1991) 83; Y. Georgelin, J.C. Wallet, Mod. Phys. Lett. A7 (1992)
1149, Phys. Rev. D50 (1994) 6610.
\bibitem{DALF}: F. Dalfovo, S. Giorgini,  L.P. Pitaevskii, S. Stringari,
Rev. Mod. Phys. 71 (1999) 463; A. J. Leggett, Rev. Mod. Phys. 73 (2001) 307.
\bibitem{FOOT1}: Our 
conventions are $\hbar$$=$$c$$=$$1$ when not explicitly written; indices $A,B$$=$$1,...,N$ label bosons; $i,j$$=$$1,2$
label coordinates in the plane, ${\bf{x}}$$\equiv$$(x_1,x_2)$,
$\epsilon_{12}$$=$$1$, 
${\bf{x}}_{AB}$$=$${\bf{x}}_A$$-$${\bf{x}}_B$; spacetime
metric is $g_{\mu\nu}$$=$diag$(+,-,-)$, $\mu,\nu,...$$=$$0,1,2$, $\epsilon_{012}$$=$$1$, 
$\partial_i^{(A)}$$\equiv$${{\partial}\over{\partial x^i_A}}$, 
$\partial_0$$\equiv$$\partial_t$, $\int_x$$\equiv$$\int
dtd{\bf{x}}$, $\int_{{\bf{x}}}$$\equiv$$\int d{\bf{x}}$, $\int_t$$\equiv$$\int dt$.
Summation over repeated indices is understood. 
\bibitem{GANG}: see e.g. D.S. Petrov, D.M. Gangardt, G.V. Shlyapnikov, J. Phys. IV France 116 (2004) 5 and references 
therein; see also M.D. Lee, S.A. Morgan, M.J. Davis, K. Burnett, Phys. Rev. A65 (2002) 043617.
\bibitem{ZHANG}: S.-C. Zhang, T.H. Hansson, S. Kivelson, Phys. Rev. Lett. 62
(1989) 82. 
\bibitem{DESER}: S. Deser, R. Jackiw, S. Templeton, Ann. Phys. 140 (1982) 372.
\bibitem{FOOOT2}: It can be realized that the general (static) solutions of
(5b) such that $f_\mu$$\ne$$-$${{1}\over{\eta}}{\cal{J}}_\mu$ have infinite
energy.
\bibitem{LATINSKY}: S. Latinski, D. Sorokin, Mod. Phys. Lett. A6 (1991) 3525.
\bibitem{F2}: ${{\Lambda}\over{\Lambda_{st}}}$$\ll$$1$ corresponds to a statistical
interaction with long range which is unlikely. This would correspond either to
the small $e^2$ regime  where MCS theory enters a strong coupling regime 
for which a description of anyons as given by (3) is questionable or to a
fractional Quantum Hall state with $\eta$$\ll$$1$, e.g. state with extremely
large denominator.
\bibitem{EZAWA}: Z.F. Ezawa, M.  Hotta, A. Iwazaki, Phys. Rev. D44 (1991) 452.
\bibitem{DENSPROF}: Inclusion of small $|\phi|^6$ terms would not alter noticeably the shape of the density profiles
discussed above; see also 2nd of \cite{FISCHER}.
\bibitem{REMARK}: For a generic Pauli 
coupling $\alpha f_\mu J^\mu$ (see $S_2$), a (static) point-like particle would produce electric and 
magnetic fields spread over a domain of size $\sim$$\Lambda_{st}$; when $\alpha$$=$$-$${{1}\over{\eta e^2}}$, 
the magnetic field becomes localized while the electric field becomes screened by Pauli coupling contributions 
so that the net electric field vanishes. 
\bibitem{PAREDES}: see B. Paredes, J.J. Garcia-Ripoll, P. Zoller, J.I. Cirac, J. Phys. IV 116 France (2004) 135 and ref. therein.

\end{thebibliography}
\end{document}